\documentclass[amsmath,10pt,twocolumn,superscriptaddress,footinbib,aip,longbibliography,reprint]{revtex4-2}

\usepackage{amsfonts}
\usepackage{amsmath}
\usepackage{mathtools}
\usepackage{siunitx}
\usepackage{graphicx}
\usepackage{bbm}
\usepackage{float}
\usepackage{algorithmicx}
\usepackage{algpseudocode}
\usepackage{algorithm}
\usepackage{color}

\DeclarePairedDelimiterX{\inner}[2]{\langle}{\rangle}{{#1},{#2}}

\usepackage[colorlinks=true,citecolor=blue,linkcolor=blue,urlcolor=blue]{hyperref}
\begin{document}

\title{Sensitivity Analysis in the Face of Rare Events}

\author{John Strahan}
\author{Todd R. Gingrich}
\affiliation{Department of Chemistry, Northwestern University,
2145 Sheridan Road, Evanston, Illinois 60208, USA}

\begin{abstract}
Molecular motors and other complex nonequilibrium systems are controlled by large sets of design parameters, and optimizing those parameters requires computing sensitivities---derivatives of dynamical observables with respect to the parameters.
When the system's dynamics involves rare events, both the observable and its sensitivity are difficult to estimate from direct simulation.
We present a practical computational pipeline that addresses both challenges by combining importance sampling with a Markov state model (MSM).
The MSM separately captures the slow, rare-event dynamics and the fast, local dynamics, and the chain rule connects those two pieces to yield an efficient sensitivity estimator.
An iterative reweighting procedure based on the RiteWeight algorithm substantially reduces approximation errors from the MSM coarse-graining.
We validate the approach on diffusion in the M\"{u}ller-Brown potential, where the sensitivity of a transition rate to landscape parameters can be computed exactly.
We then use sensitivies to optimize the directional bias of a particle-based model of a catalysis-driven molecular motor.
\end{abstract}

\maketitle

\section{Introduction}

Many models of complex nonequilibrium systems are controlled by a large number of parameters, and optimizing performance requires sensitivities\textemdash derivatives of dynamical observables with respect to those parameters.
One might use gradient descent to find a time-varying protocol that drives self-assembling components into a desired product in finite time~\cite{goodrich_designing_2021,trubiano_optimization_2022}.
Alternatively, one could tune the interaction strengths between motifs of a chemically fueled molecular motor to increase its directional bias~\cite{albaugh_simulating_2022}.
In either case, each step of the optimization requires an efficient way to compute sensitivities of the relevant dynamical observable with respect to many parameters simultaneously.

Recent years have seen rapid progress in computing such sensitivities for molecular systems.
Automatic differentiation through molecular dynamics trajectories can yield gradients of thermodynamic and kinetic observables with respect to all force-field parameters simultaneously~\cite{greener_differentiable_2021,wang_differentiable_2020,schoenholz_jax_2020,greener_reversible_2025}, and the same technique has been applied to optimize nonequilibrium control protocols~\cite{engel_optimal_2023} and to fit coarse-grained models of nucleic acids to experimental data~\cite{krueger_differentiable_2024}.
Likelihood ratio (score function) estimators~\cite{glynn_likelihood_1990} and pathwise derivative methods~\cite{rathinam_efficient_2010,sheppard_pathwise_2012,pantazis_parametric_2013,anderson_efficient_2012} offer complementary routes that do not require differentiating through the integrator but instead reweight or perturb trajectories.

These methods work well when trajectories are long enough to observe the events that determine the observable.
The deeper obstacle is rare events.
When the observable is controlled by infrequent barrier crossings, a finite simulation may observe none of them.
No estimator can extract a gradient signal that the simulation does not contain.
Markov state models were developed precisely to address this timescale separation~\cite{husic_markov_2018,prinz_markov_2011,schutte_direct_1999}.
By coarse-graining the state space and building a transition matrix from short trajectory fragments, an MSM captures the slow dynamics without requiring any single trajectory to span the rare crossings.
Related Galerkin approximation schemes for dynamical observables share the same timescale-separation strategy~\cite{thiede_galerkin_2019,lorpaiboon_augmented_2022}.

Differentiating through coarse-grained kinetic models provides a more direct route to the rare-event regime.
Indeed, \citet{trubiano_optimization_2022} combined MSM analysis with adjoint-based gradients to optimize time-dependent self-assembly protocols, exploiting the fact that the sensitivity of a Markov chain's stationary distribution to its transition matrix is a well-studied algebraic problem~\cite{meyer_sensitivity_1994,thiede_sharp_2015}.
That MSM-based approach, however, operated in a regime where the MSM could be built from direct simulation and the gradient of interest was the response of a finite-horizon yield to a time-dependent protocol.
In the rare-event regime, the stationary distribution itself must be inferred from reweighted data, and the sensitivity requires differentiating not just through the transition matrix but through the importance weights that reconstruct the stationary measure from short trajectory fragments.

What has been missing is a pipeline suited to the rare-event regime, where the stationary distribution cannot be sampled directly and must instead be inferred.
We describe such a pipeline here.
We adapt the recently introduced RiteWeight algorithm \cite{kania_riteweight_2025}, which iteratively refines importance weights for steady-state estimation, to simultaneously track the derivative of those weights with respect to model parameters.
The result is a practical sensitivity estimator that maintains performance as the temperature decreases and rare transitions become exponentially less frequent.
We validate the method on the M\"{u}ller-Brown potential \cite{muller_location_1979}, a two-dimensional landscape with well-separated metastable states whose exact sensitivities are available for comparison, and show that accuracy is maintained as the temperature is lowered.
We then apply the MSM-based sensitivity pipeline to the particle-based catenane motor model of \citet{albaugh_simulating_2022}, demonstrating that gradient descent on the interaction parameters progressively improves the motor's directional bias.

The manuscript proceeds as follows.
Section~\ref{sec:sec2} establishes the sensitivity formula, decomposing the derivative of a reweighted dynamical expectation into a propagator derivative, computable from trajectory data, and an importance weight derivative obtained from the algebraic sensitivity of the MSM stationary distribution.
Section~\ref{sec:sec3} develops the RiteWeight adaptation for carrying sensitivity derivatives through the iterative reweighting.
Section~\ref{sec:sec4} validates the method on the M\"{u}ller-Brown potential across temperatures, bin counts, and lag times.
Section~\ref{sec:sec5} applies the method to optimize the molecular motor.

\section{The Sensitivity of a Dynamical Order Parameter}
\label{sec:sec2}

This section derives the working formula for the sensitivity of a steady-state observable to a model parameter.
The key step is a decomposition of the derivative into two pieces with very different characters.
One piece involves only single-step transition probabilities: how does each individual step of the dynamics change when the parameter is varied?
That piece is straightforward to estimate from trajectory data.
Because the MSM decomposition only evaluates it on short trajectory fragments, its variance stays controlled even as the rare-event timescale grows.
The other piece involves the stationary distribution itself: how does the long-time probability of being in each part of state space change with the parameter?
That piece is the hard one.
The stationary distribution is the eigenvector of the dynamics, and there is generally no closed-form expression for how it responds to a perturbation.
The rest of this section shows how to handle both pieces.
For the first, we differentiate the integrator's transition kernel explicitly.
For the second, we introduce importance weights whose derivatives can be estimated via a Markov state model.

Let $Z_t$ be a Markov process with transition kernel $P_{\lambda}$,
\begin{equation}
    \mathbbm{P}_{\lambda}[Z_{dt}=y \mid Z_0=x]=P_\lambda(x,y),
\end{equation}
depending smoothly on a parameter $\lambda$.
We denote the stationary expectation of a dynamical observable $F(Z_0,\ldots,Z_t)$ by
\begin{equation}\label{eq:orderparam}
    \langle F \rangle_{\pi_\lambda} \equiv \int F(z_0,\ldots,z_t)\,\pi_\lambda(z_0)\,dz_0\prod_{s=1}^{t}P_\lambda(z_{s-1},z_s)\,dz_s,
\end{equation}
where $\pi_\lambda$ is the stationary distribution of $P_\lambda$.
The form is quite general.
$F$ could be an instantaneous positional observable, a configurational order parameter, or a path observable like a transition rate or molecular motor's directional bias.

Differentiating Eq.~\eqref{eq:orderparam} with respect to $\lambda$ yields the two-term decomposition anticipated above:
\begin{equation}
  \begin{split}
  &\frac{d\langle F\rangle_{\pi_\lambda}}{d\lambda}
  = \left\langle F\,\frac{d \log \pi_{\lambda}(Z_0)}{d\lambda}\right\rangle_{\pi_\lambda} \\
  &\quad+ \left\langle F\,\frac{d}{d\lambda}\log \bigl[P_\lambda(Z_0, Z_1) \cdots P_\lambda(Z_{t-1},Z_t)\bigr]\right\rangle_{\pi_\lambda}.
  \end{split}
  \label{eq:decomposition}
\end{equation}
The second term is the log-derivative of the path probability.
As the sum of single-step log-derivatives, it is directly computable from trajectory data for any model with an explicit transition kernel.
The same quantity appears as the reweighting factor when parameters are perturbed within transition path sampling \cite{dellago_transition_2002}.
It is known as the score function or REINFORCE estimator in the stochastic optimization literature \cite{glynn_likelihood_1990,williams_simple_1992} and as the Girsanov weight in the theory of stochastic processes \cite{oksendal_stochastic_2003}.
The first term of Eq.~\eqref{eq:decomposition} is the hard one.
It requires $\pi_\lambda$ (the top eigenvector of the Markov process) and its derivative with respect to $\lambda$.
We are focused on situations in which the state space is sufficiently large (often continuous) that it is impractical to explicitly compute $\pi_\lambda$, much less its sensitivity to $\lambda$.
For discrete-state Markov jump processes, exact identities based on the matrix tree theorem and related graph-theoretic tools can bound or express these sensitivities without enumerating the full state space \cite{owen_universal_2020,fernandes_martins_topologically_2023,aslyamov_nonequilibrium_2024,aslyamov_general_2024,zheng_universal_2024}.
For continuous-state diffusions, information-theoretic methods based on R\'enyi divergences provide rigorous bounds on rare-event sensitivities directly from path-space relative entropies~\cite{dupuis_sensitivity_2020}.
Our approach is more computational in nature; we want to estimate the sensitivity with a sampling procedure.

To make the stationary distribution derivative tractable, it is useful to rewrite Eq.~\eqref{eq:orderparam} as an expectation not over $\pi_\lambda$ but over a fixed reference distribution $\mu$.
Let $\mu$ be any distribution whose support contains that of $\pi_\lambda$, and define $w_\lambda(x) = \pi_\lambda(x)/\mu(x)$ as importance weights that encode all the $\lambda$-dependence of the stationary measure.
Then
\begin{equation}
    \langle F \rangle_{\pi_\lambda} = \langle w_\lambda\, F \rangle_{\mu},
\end{equation}
and differentiating both sides gives
\begin{equation}\label{eq:ExpDeriv}
  \begin{split}
    &\frac{d\langle F\rangle_{\pi_\lambda}}{d\lambda}
    = \left\langle F\,\frac{d w_\lambda(Z_0)}{d\lambda}\right\rangle_{\mu} \\
    &\quad+ \left\langle w_\lambda\,F\,\frac{d}{d\lambda}\log\bigl[P_{\lambda}(Z_0,Z_1)\cdots P_{\lambda}(Z_{t-1},Z_t)\bigr]\right\rangle_{\mu}.
  \end{split}
\end{equation}
The structure of the two terms is the same as before, but the troublesome $d\log\pi_\lambda/d\lambda$ has been replaced by $dw_\lambda/d\lambda$, the derivative of the importance weights.
The advantage of this rewriting is twofold.
First, the reference distribution $\mu$ can be chosen to have good coverage of the full state space---including rare-event regions---regardless of whether those regions are heavily weighted by $\pi_\lambda$.
By sampling trajectory starts from $\mu$ (for instance, uniformly over a region of interest) rather than from $\pi_\lambda$ directly, one can populate rare regions without waiting for long spontaneous trajectories to visit them.
Second, because $w_\lambda = \pi_\lambda/\mu$ is a ratio of probability densities, it can be approximated by a Markov state model at the coarse level.
The MSM need only estimate the relative probability that the stationary measure assigns to each coarse cell, a much more tractable target than computing $d\log\pi_\lambda/d\lambda$ directly.

\subsection{Computing the Importance Weight Derivatives via an MSM}
\label{sec:sec2a}

The key observation is that $w_\lambda(x) = \pi_\lambda(x)/\mu(x)$ need not be estimated pointwise across the continuous space of microstates $x$.
If we partition microstates into coarse cells $\{S_j\}$, we can instead approximate $w_{\lambda}(x)$ as a piecewise constant function taking the value $w_{\lambda, j}$ for all $x \in S_{j}$.
Those weights for each cell are approximated as $c_j / \mu_{j}$, where now $c_j$ and $\mu_j$ are the probabilities of occupying cell $j$ in the stationary and reference distributions, respectively.
Procedurally, one chooses $\mu$, which sets the fraction of trajectories initialized in cell $j$: $\mu_j$.
From those trajectories, one infers a Markov state transition matrix and the $j^{\rm th}$ component of the stationary vector for this transition matrix is the estimate for $\pi_{\lambda}$ in cell $j$: $c_j$.

More formally, let $N$ trajectory fragments be initialized in $\mu$ at time $0$ and propagated to time $t$: $\{(Z^i_0,\ldots,Z^i_t)\}_{i=1}^N$.
Let $J^i_s$ denote the coarse cell index of trajectory number $i$ at time $s$, i.e.\ $Z_s^i \in S_{J_s^i}$.
The coarse transition matrix is estimated as
\begin{equation}
    P_{ab} = \frac{\sum_i \mathbbm{1}_{S_a}(Z^i_0)\,\mathbbm{1}_{S_b}(Z^i_t)}
                  {\sum_i \mathbbm{1}_{S_a}(Z^i_0)},
\end{equation}
where $\mathbbm{1}_{S_a}(x)$ is the indicator function, equal to 1 when $x \in S_a$ and 0 otherwise.
The number of coarse states is sufficiently modest that it is straightforward to numerically compute the top eigenvector of that transition matrix, which satisfies $\sum_j c_j P_{jk} = c_k$.
The estimated importance weight contributed by trajectory $i$ is
\begin{equation}
  w_\lambda(Z^i_0) = \frac{c_{J^i_0}}{\sum_\ell \mathbbm{1}_{S_{J^i_0}}(Z^\ell_0)}.
\end{equation}

Since $w_\lambda$ depends on $\lambda$ only through the coarse stationary distribution $c$ and the transition matrix $P$, the chain rule gives
\begin{equation}\label{eq:ChainRule}
    \frac{d w_\lambda(Z^i_0)}{d\lambda}
    = \frac{1}{\sum_\ell \mathbbm{1}_{S_{J^i_0}}(Z^\ell_0)}
      \sum_{pq}\frac{d c_{J^i_0}}{d P_{pq}}\frac{d P_{pq}}{d\lambda}.
\end{equation}
The two factors reflect the two timescales of the problem.
The first factor, $dc_k/dP_{pq}$, encodes how the long-time coarse-grained populations respond to a change in a single transition matrix element.
Because the coarse chain has only $M$ states, this is a finite linear algebra problem.
We use the perturbation formula of \citet{thiede_sharp_2015},
\begin{equation}\label{eq:PDerv}
    \frac{d c_{k}}{d P_{pq}}
    = c_{p}\!\left(\mathbbm{E}_{J_0=q}\!\left[\sum_{s=0}^{\tau_{p}-1}\mathbbm{1}_{J_s=k}\right] - c_{k}\,\mathbbm{E}_{J_0=q}[\tau_{p}]\right),
\end{equation}
where $\mathbbm{E}_{J_0=q}$ denotes the expectation conditional on the coarse chain starting in state $q$ and $\tau_p = \min\{s>0 : J_s = p\}$ is the first return time to state $p$.
The factor in parentheses measures how much time the chain started at $q$ spends in state $k$ before returning to $p$, relative to the expected return time weighted by $c_k$---a mean-first-passage-time calculation that reduces to a linear system.
Specifically, defining $F_{qpk} = \mathbbm{E}_{J_0=q}[\sum_{s=0}^{\tau_p-1}\mathbbm{1}_{J_s=k}]$ as the expected number of visits to $k$ before the chain first returns to $p$, for fixed $p$ and $k$ this satisfies the Feynman-Kac system
\begin{equation}
    \sum_{r \neq p}(P_{qr} - I_{qr}) F_{rpk} = \delta_{qk}, \qquad F_{ppk} = 0,
\end{equation}
where the linear system is solved over $r,q\neq p$ with $p$ acting as an absorbing index, and the boundary $F_{ppk}=0$ encodes the convention that no visits are accumulated once the chain has reached~$p$.
The system is solved for all $(p,k)$ pairs by standard linear solvers, with the mean return time following as $\mathbbm{E}_{J_0=q}[\tau_p] = \sum_k F_{qpk}$.
The second factor, $dP_{ab}/d\lambda$, encodes how the short-time, local transition probabilities between coarse cells change with the parameter.
We compute it by applying Eq.~\eqref{eq:ExpDeriv} with the indicator observable $F(Z^i_0,\ldots,Z^i_t) = \mathbbm{1}_{S_a}(Z^i_0)\,\mathbbm{1}_{S_b}(Z^i_t)$, recovering $dP_{ab}/d\lambda$ from the same trajectory data used for everything else.

\subsection{Derivative of the Propagator}
\label{sec:sec2b}

The second term in Eq.~\eqref{eq:ExpDeriv} requires the log-derivative of the transition kernel, $d\log P_\lambda/d\lambda$, summed over the steps of each trajectory.
For any integrator with an explicit transition density, this derivative can be computed analytically from the noise realization at each step.
The concrete form depends on the integrator.
For overdamped Brownian dynamics discretized with the Euler-Maruyama scheme, the propagator is Gaussian in the positions and the log-derivative is straightforward (Section~\ref{sec:sec4}).
For Langevin dynamics with momenta, the choice of splitting scheme matters.
We use the ABOBA integrator, whose propagator has a closed Gaussian form that makes its log-derivative analytically tractable, and derive the explicit expression in Section~\ref{sec:sec5a_integrator}.

\section{Reducing Approximation Error with RiteWeight}
\label{sec:sec3}

The importance weight formula of Section~\ref{sec:sec2a} approximates the true change of measure by a piecewise-constant function defined by the coarse partition.
If the partition is coarse, the approximation can be poor; if it is fine, the MSM may not be well-estimated from the available data.
The recently introduced RiteWeight algorithm \cite{kania_riteweight_2025} iteratively reduces this approximation error without requiring a finer partition.
Here we describe how to adapt RiteWeight to track not just the weights but also their derivatives with respect to $\lambda$.

The idea is simple.
At each iteration, we use the current weights to build a new MSM, extract a new coarse stationary distribution, and use it to correct the weights.
If the initial coarse partition was too coarse to resolve the stationary distribution accurately, the corrected weights will better represent the true measure, and the next MSM built from those corrected weights will be more accurate still.
The derivative of the weights can be carried through each iteration by the product rule, so the sensitivity estimate improves alongside the weights themselves.

At iteration $\alpha$, we have weights $\{w^\alpha_\lambda(Z^i_0)\}$ and a coarse partition $\{S^\alpha_j\}$.
The reweighted coarse transition matrix is
\begin{equation}
    P^\alpha_{ab} = \frac{\sum_i w^\alpha_\lambda(Z^i_0)\,\mathbbm{1}_{S^\alpha_a}(Z^i_0)\,\mathbbm{1}_{S^\alpha_b}(Z^i_t)}
                        {\sum_i w^\alpha_\lambda(Z^i_0)\,\mathbbm{1}_{S^\alpha_a}(Z^i_0)},
\end{equation}
with stationary distribution $c^\alpha$ satisfying $\sum_a c^\alpha_a P^\alpha_{ab} = c^\alpha_b$.
The weights are then updated as
\begin{equation}
    w^{\alpha+1}(Z^i_0)
    = (1-\eta^\alpha)w^\alpha(Z^i_0)
    + \eta^\alpha \frac{w^\alpha(Z^i_0)\,c^\alpha_{J^i_0}}
                       {\sum_\ell w^\alpha(Z^\ell_0)\,\mathbbm{1}_{S^\alpha_{J^i_0}}(Z^\ell_0)},
\end{equation}
where $\eta^\alpha \in (0,1]$ is a learning rate.
Taking the derivative of this update with respect to $\lambda$ gives the companion recursion for $dw^\alpha_\lambda/d\lambda$:
\begin{multline}
    \frac{d w_\lambda^{\alpha+1}(Z^i_0)}{d\lambda}
    = (1-\eta^\alpha)\frac{d w_\lambda^\alpha(Z^i_0)}{d\lambda}
    \\ + \eta^\alpha\frac{d w_\lambda^\alpha(Z^i_0)}{d\lambda}
      \frac{c^\alpha_{J^i_0}}{\sum_\ell w^\alpha(Z^\ell_0)\,\mathbbm{1}_{S^\alpha_{J^i_0}}(Z^\ell_0)} \\
    + \frac{\eta^\alpha\,w^\alpha(Z^i_0)}{D^\alpha_i}
      \sum_{ab}\frac{d c^\alpha_{J^i_0}}{dP^\alpha_{ab}}\frac{dP^\alpha_{ab}}{d\lambda} \\
    - \frac{\eta^\alpha\,w^\alpha(Z^i_0)\,c^\alpha_{J^i_0}}{(D^\alpha_i)^2}
      \sum_\ell \frac{d w^\alpha(Z^\ell_0)}{d\lambda}\,\mathbbm{1}_{S^\alpha_{J^i_0}}(Z^\ell_0).
\end{multline}
Here $D^\alpha_i = \sum_\ell w^\alpha(Z^\ell_0)\,\mathbbm{1}_{S^\alpha_{J^i_0}}(Z^\ell_0)$ is the reweighted count in the coarse state containing $Z^i_0$. The coarse stationary distribution derivative $dc^\alpha_{J^i_0}/dP^\alpha_{ab}$ is computed via Eq.~\eqref{eq:PDerv} with the reweighted transition matrix, and $dP^\alpha_{ab}/d\lambda$ is estimated as in Section~\ref{sec:sec2a} using the current weights.

The four terms have a clear structure.
The first two terms propagate the existing weight derivative forward, modulated by the new coarse stationary distribution---they are the counterpart, for the derivative, of the weight update itself.
The third term is the genuinely new piece.
It accounts for how the coarse stationary distribution changes with $\lambda$ at this iteration, injecting fresh sensitivity information from the current MSM.
The fourth term corrects for the fact that the normalization denominator $D^\alpha_i$ itself depends on the weights and therefore on $\lambda$.
Each RiteWeight iteration therefore carries along a sensitivity estimate at no fundamental change in algorithmic complexity.

In practice, the partition $\{S^\alpha_j\}$ is redrawn at each iteration (for instance, by choosing new Voronoi centers from the dataset), so successive iterations probe the state space at different resolutions.
The learning rate $\eta^\alpha$ controls how aggressively the weights are updated.
We find that a few iterations with $\eta^\alpha = 1$ suffice for the problems studied here, consistent with the findings of \citet{kania_riteweight_2025}.

\section{M\"{u}ller-Brown Potential}
\label{sec:sec4}

We first validate the method on the M\"{u}ller-Brown (MB) potential \cite{muller_location_1979}:
\begin{multline}
    \label{eq:MB}
V_{\rm MB}(y, z)=\frac{1}{20}\sum_{i=1}^4 C_i \exp\!\left[a_i(y-y_i)^2+\right. \\ \left. b_i(y-y_i)(z-z_i)+c_i(z-z_i)^2\right],
\end{multline}
a canonical two-dimensional landscape with well-separated metastable states connected by a saddle point.
The MB potential is small enough to admit an exact treatment.
The transition rate $k_{AB}$ and its sensitivity to a landscape parameter can be computed on a discrete spatial grid by direct matrix methods, with no sampling noise.
This exact check is unavailable for the high-dimensional motor of Sec.~\ref{sec:sec5}, whose configuration space spans the coordinates of dozens of interacting particles.
MB therefore provides a rare opportunity to verify the method against ground truth before deploying it where no such check exists.

\begin{figure}[bt]
\begin{center}
\includegraphics[width=.45\textwidth]{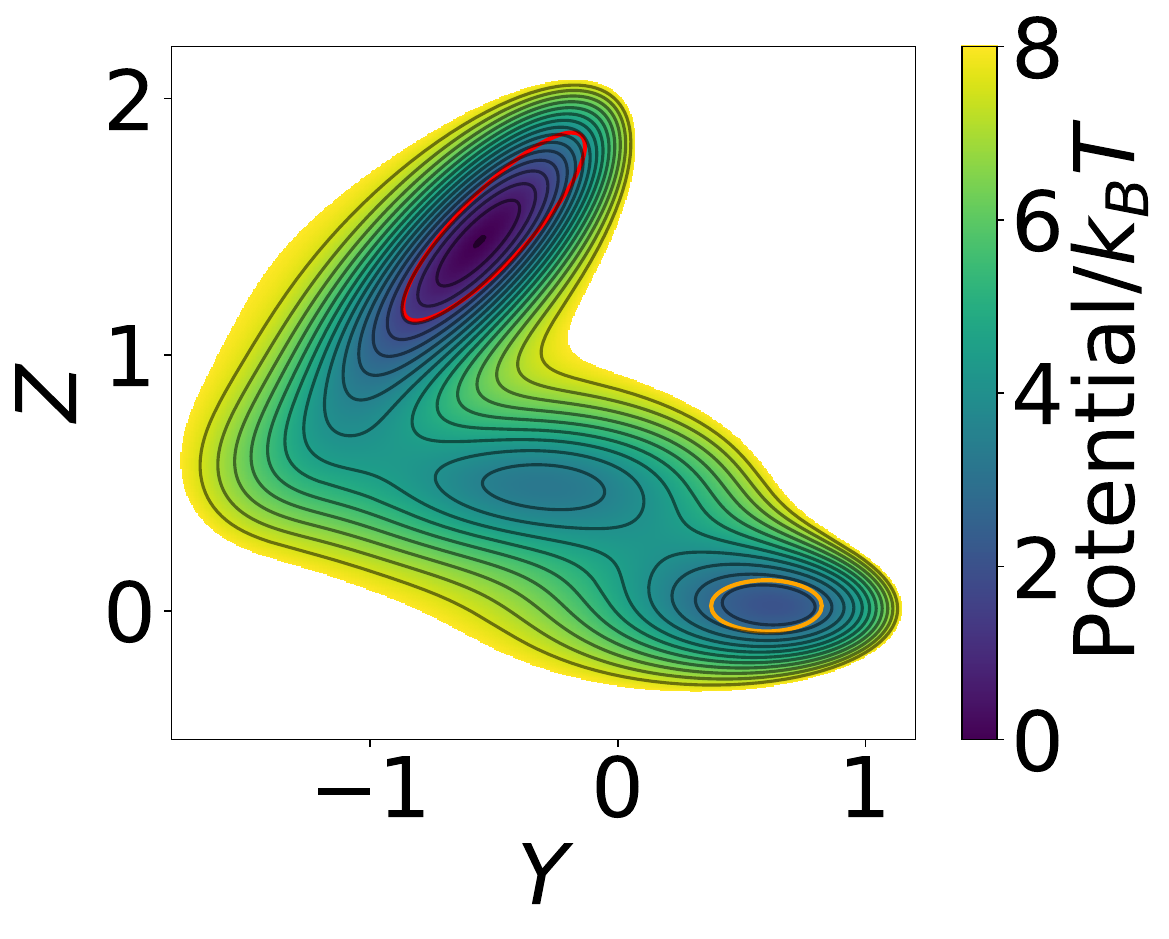}
\end{center}
\caption{\label{fig:MB}
M\"{u}ller-Brown potential energy landscape, used here as a validation testbed.
The potential has three local minima, two deep metastable states $A$ (upper left, orange ellipse) and $B$ (lower right, red ellipse), and a shallower intermediate basin between them whose depth is controlled by the parameter $C_2$.
At low temperature, transitions between $A$ and $B$ are rare.
We test whether the method correctly estimates the sensitivity $dk_{AB}^{-1}/dC_2$---how the $A\to B$ rate responds to changes in the intermediate's energy---against exact numerical reference values.
Contours are spaced every $0.5\,k_{\rm B}T$.
}
\end{figure}

Figure~\ref{fig:MB} shows the potential, with the two metastable states $A$ and $B$ highlighted.
The parameters are set to
\begin{equation}
\begin{aligned}
(C_1,C_2,C_3,C_4) &= (-200,-100,-170,15),\\
(a_1,a_2,a_3,a_4) &= (-1,-1,-6.5,0.7),\\
(b_1,b_2,b_3,b_4) &= (0,0,11,0.6),\\
(c_1,c_2,c_3,c_4) &= (-10,-10,-6.5,0.7),\\
(y_1,y_2,y_3,y_4) &= (1,-0.27,-0.5,-1),\\
(z_1,z_2,z_3,z_4) &= (0,0.5,1.5,1).
\end{aligned}
\end{equation}
The dynamics of the position $X_t = (y,z)_t$ are overdamped Brownian motion discretized with the Euler-Maruyama scheme,
\begin{equation}
    X^j_{t+1} = X^j_t - dt\,\nabla V_\text{MB}(X_t) + \sqrt{2\,dt\,\beta^{-1}}\;\xi^j_t,
\end{equation}
where $\beta = 1/k_{\rm B}T$ is the inverse temperature and $\xi^j_t$ is a standard normal noise.
Transitions between $A$ and $B$ constitute the rare events that challenge direct sensitivity estimation.
At low temperature, the mean crossing time grows exponentially with the barrier height, so any computationally feasible ensemble of trajectories will observe few or no $A\to B$ events.
An estimator built from such an ensemble cannot reliably extract $k_{AB}$, let alone its sensitivity to a landscape parameter.

The parameters of Eq.~\eqref{eq:MB} are the standard Müller-Brown parameters \cite{muller_location_1979} except that $y_2$ has been shifted from $0$ to $-0.27$, which deepens the intermediate basin near $(-0.5, 1.5)$.
This modification is deliberate.
Simply lowering the temperature on the unmodified landscape makes barrier crossings rarer but does not make the problem structurally harder, since the kinetics remain effectively two-state.
With the deepened intermediate, lowering the temperature has a qualitatively different effect.
Trajectories transitioning from $A$ to $B$ must pass through the intermediate basin, which itself becomes a kinetic trap at low temperature.
The dwell time in the intermediate grows exponentially as the temperature decreases, so the coarse-grained model must correctly resolve three metastable regions---not merely two---and accurately capture the populations and fluxes among them.
The sensitivity to $C_2$, the depth of this intermediate basin, is therefore a particularly demanding test.
An accurate estimate requires the method to faithfully represent how changes to the intermediate affect the overall $A \to B$ rate.

To compute how $k_{AB}$ changes with a landscape parameter such as $C_2$, we must cast the rate as an expectation of the form~\eqref{eq:orderparam}.
This requires a stationary distribution $\pi$ and a path functional $F$ that depends on only a short trajectory segment.
A naive choice, taking $\pi$ to be the Boltzmann distribution and $F$ to count $A\!\to\!B$ transitions, fails because the rate is defined through a long-time limit, forcing trajectories long enough to capture the rare crossing.
The likelihood ratio estimator for the second term of Eq.~\eqref{eq:decomposition} then becomes very noisy, since the variance of the score-function sum scales with the trajectory length~\cite{arampatzis_efficient_2016}.
Centering strategies can bound this variance for generic steady-state averages, but a cleaner solution is to augment the state space with a label recording whether the system was last in $A$ or in $B$.
In this enlarged space, the transition rate becomes an instantaneous current from the ``last in $A$'' state to the ``last in $B$'' state, and $F$ need only examine a single timestep.
We make this construction precise below and then write $k_{AB}$ as a ratio of two expectations of the form Eq.~\eqref{eq:orderparam}.

The augmented Markov process is $Z_t=(X_t,\mathbbm{1}_A(X_{T^-_{A\cup B}}))$, where $T^-_{A\cup B}=\min\{t>0:X_{-t} \in A \cup B\}$ records the time of the most recent visit to $A$ or $B$.
The augmented variable is simply a binary label.
It equals 1 if the system was last in $A$ and 0 if it was last in $B$.
The $A\!\to\!B$ transition rate is the instantaneous flux out of the ``last in $A$'' population,
\begin{equation}\label{eq:kAB}
    k_{AB} = \lim_{t\to 0}\frac{1}{t}\frac{\langle\mathbbm{1}_A(X_{T^-_{A\cup B}})\,\mathbbm{1}_B(X_{t})\rangle_{\pi}}{\langle\mathbbm{1}_A(X_{T^-_{A\cup B}})\rangle_{\pi}},
\end{equation}
where $\pi$ now denotes the stationary distribution of the augmented process.  This is the standard reactive flux expression for the transition rate~\cite{chandler_statistical_1978,e_transition-path_2010}.
The numerator measures the probability that a trajectory last in $A$ reaches $B$ within time $t$.
The prefactor $1/t$ converts this to a flux.
The denominator is the fraction of time the system spends having last visited $A$.

Equation~\eqref{eq:kAB} is a ratio of two expectations over $\pi$, each of the form treated in Eq.~\eqref{eq:orderparam}.
We apply Eq.~\eqref{eq:ExpDeriv} to compute the sensitivity of both the numerator and denominator to a landscape parameter, then combine using the quotient rule to get the sensitivity of $k_{AB}$.
Both the numerator and denominator depend on $\lambda$ through the same stationary distribution $\pi_\lambda$, so the importance weights and their MSM derivatives need only be computed once.
The same is true of the propagator log-derivative, which depends on the integrator and the landscape, not on the choice of observable $F$.
For the Euler-Maruyama scheme used here, the propagator is Gaussian and the log-derivative takes the form
\begin{equation}\label{eq:DerivEM}
    \frac{d \log P_{\lambda}(X_{t+1}\mid X_t)}{d\lambda}
    = -\sqrt{\frac{dt\,\beta}{2}}\sum_j \xi^j_t \,\frac{d(\nabla V_\lambda)^j(X_t)}{d\lambda}.
\end{equation}

The exact rate is computed on a discrete spatial grid using the scheme of \citet{lorpaiboon_augmented_2022}.
The sensitivity $dk_{AB}^{-1}/dC_2$ is obtained by central finite differences of the exact inverse rate,
\begin{equation}\label{eq:FD}
    \frac{dk_{AB}^{-1}}{dC_2} \approx \frac{k_{AB}^{-1}(C_2+\epsilon)-k_{AB}^{-1}(C_2-\epsilon)}{2\epsilon},
\end{equation}
with $\epsilon = 10^{-4}$.
To test the method, we sample initial conditions uniformly from the region $V_\text{MB}(x) < 7.0$, assign last-visited-state labels, run a short burn-in of $0.1$ time units, and propagate each starting condition for a lag time $t$.
We then build a Voronoi partition of the state space by selecting $M$ cluster centers from the dataset via $k$-means clustering.

The lag time $t$ requires care.
The reactive flux expression Eq.~\eqref{eq:kAB} involves a $t\to 0$ limit, but that limit cannot be taken literally with finite data.
If $t$ is too small, biases from the non-Markovianity of the coarse-grained process make the MSM stationary distribution poorly determined.
If $t$ is too large, on the order of $1/k_{AB}$ or longer, trajectory fragments are long enough to capture the rare transition itself, defeating the purpose of the MSM decomposition.
A good lag time is large enough for trajectories to explore their local cell but still much shorter than $1/k_{AB}$.  At $\beta=2.0$, $1/k_{AB}$ exceeds our chosen lag time $t=0.5$ by roughly three orders of magnitude.

\begin{figure}[th]
\begin{center}
\includegraphics[width=.45\textwidth]{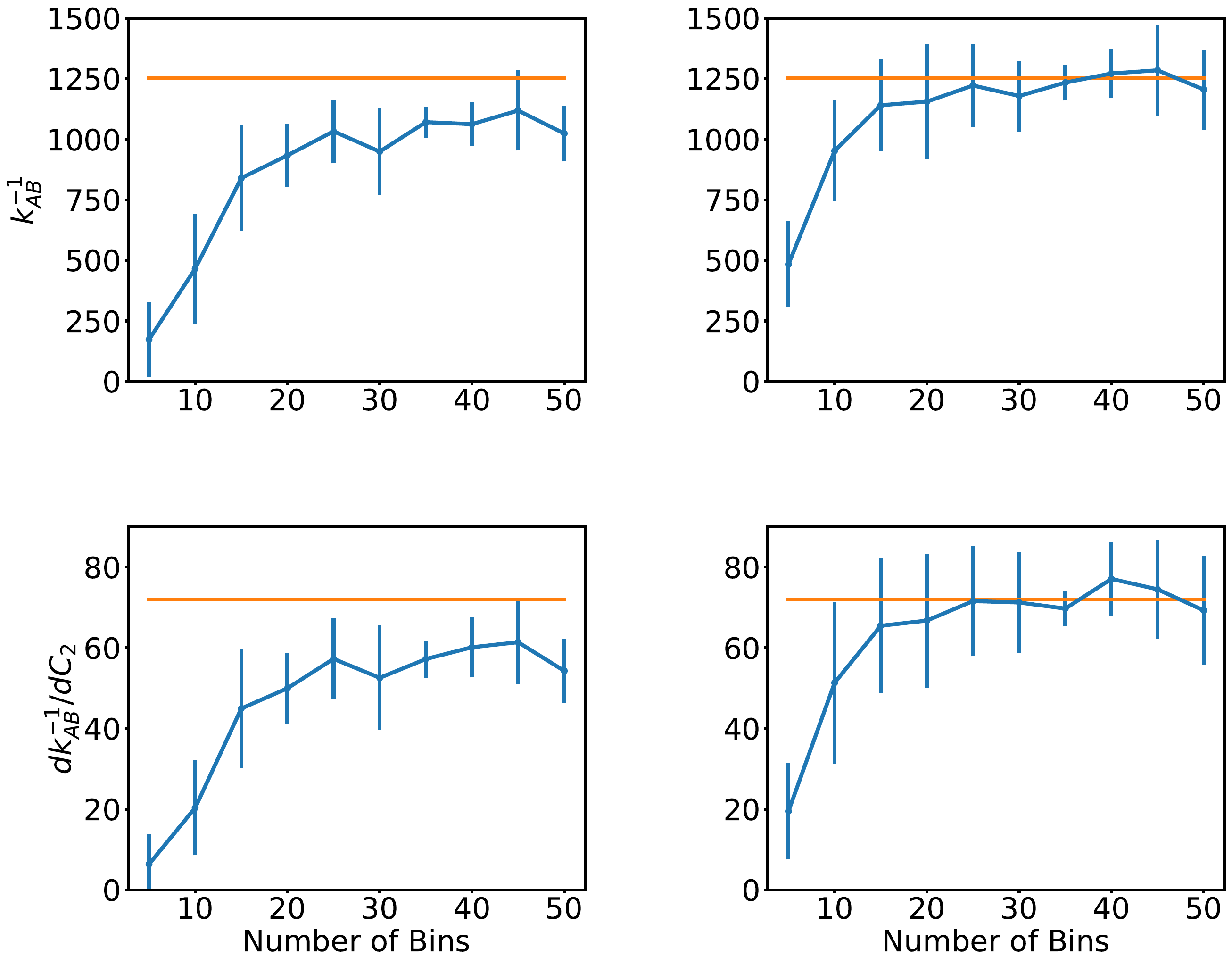}
\end{center}
\caption{\label{fig:MB_Nbin}
Sensitivity of the M\"{u}ller-Brown transition rate to the landscape parameter $C_2$, as the number of Voronoi cells $M$ is varied.
Top row: estimated inverse rate $1/k_{AB}$ from a single-iteration MSM (left) and from RiteWeight (right).
Bottom row: estimated sensitivity $dk_{AB}^{-1}/dC_2$ from an MSM (left) and from RiteWeight (right).
Horizontal dashed lines indicate the numerically exact reference.
Lag time $t = 0.5$, inverse temperature $\beta = 2.0$.
}
\end{figure}

\begin{figure}[th]
\begin{center}
\includegraphics[width=.45\textwidth]{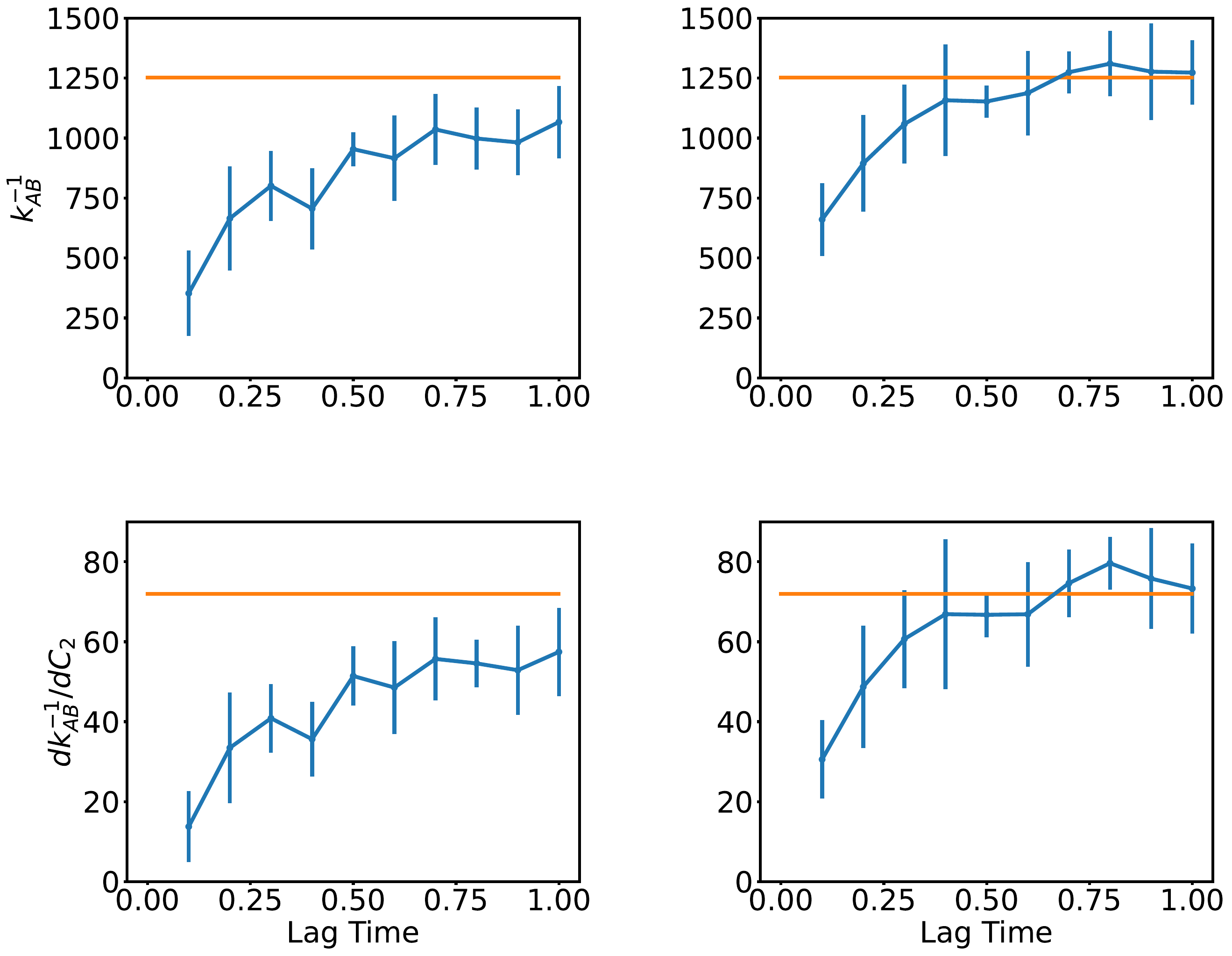}
\end{center}
\caption{\label{fig:MB_Tau}
Same layout as Fig.~\ref{fig:MB_Nbin}, with $M = 20$ fixed and the lag time $t$ varied.
Both the rate and sensitivity estimates stabilize once $t$ is large enough for trajectories to explore their local cell, and remain accurate well below $1/k_{AB}$.
Inverse temperature $\beta = 2.0$, $N = 2\times10^5$ fragments.
}
\end{figure}

\begin{figure}[th]
\begin{center}
\includegraphics[width=.45\textwidth]{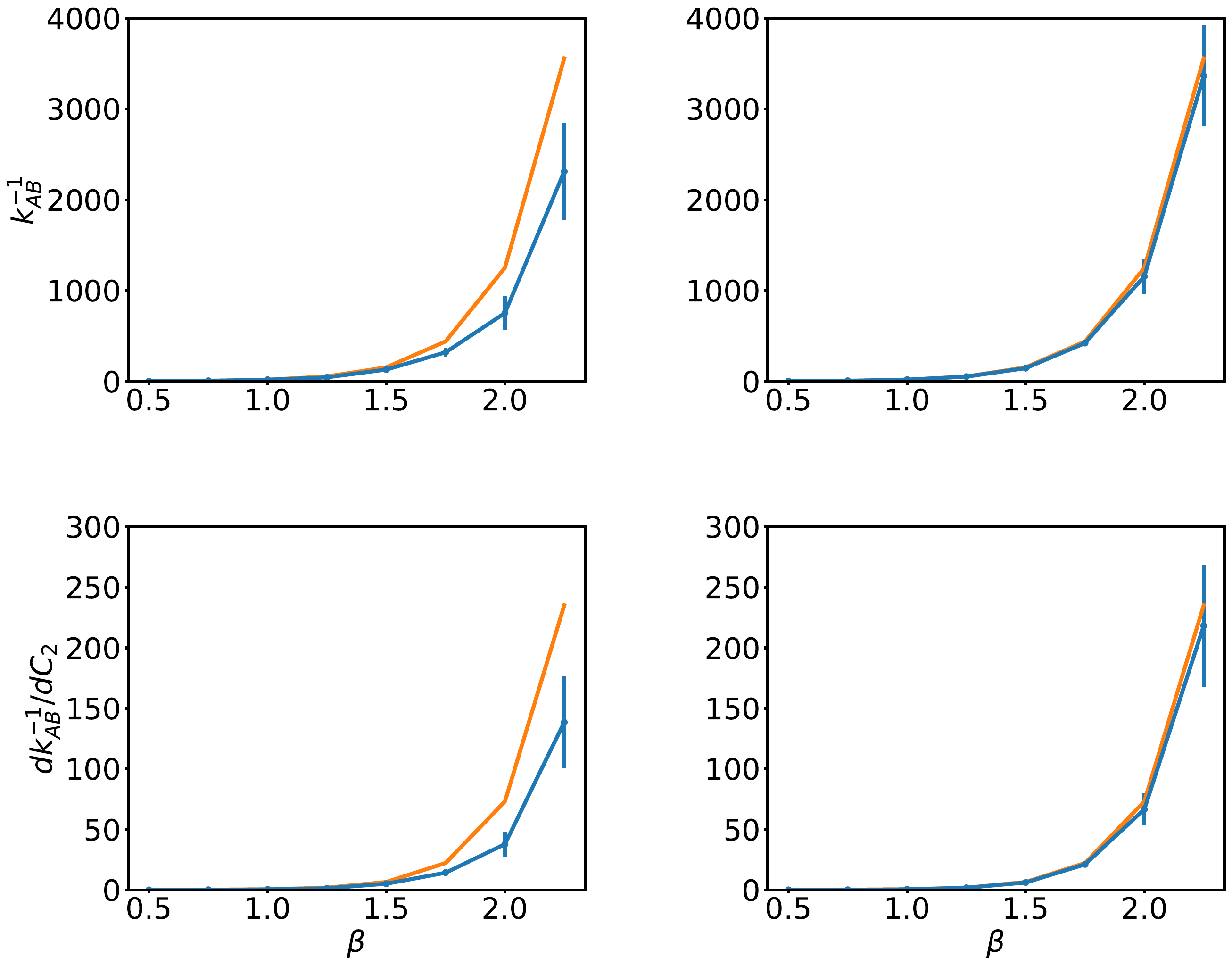}
\end{center}
\caption{\label{fig:MB_Beta}
Performance as the rare-event regime deepens.
Inverse rate and sensitivity as the inverse temperature $\beta$ is increased from 2 to 6, with $M = 20$ and $t = 0.5$ fixed.
The number of trajectory fragments is scaled as $N = 5\times10^4\,\beta^2$ to hold the statistical error roughly constant across temperatures (see text).
RiteWeight (right column) maintains accuracy throughout; the single-iteration MSM (left column) shows growing error as $\beta$ increases.
}
\end{figure}

We examine how the estimates depend on the number of Voronoi cells $M$, the lag time $t$, and the inverse temperature $\beta$, comparing in each case the single-iteration MSM estimate against the RiteWeight estimate.
Figure~\ref{fig:MB_Nbin} shows the inverse rate and sensitivity to $C_2$ at $\beta = 2$, using $N = 2\times10^5$ fragments each of length $t = 0.5$, as the number of Voronoi cells $M$ is varied.
The RiteWeight estimate converges to the exact reference once the cell count exceeds roughly 15, while the single-iteration MSM retains substantial bias across all tested cell counts.
Figure~\ref{fig:MB_Tau} holds $M = 20$ and $N = 2\times10^5$ fixed while sweeping the lag time $t$.
The estimates stabilize once $t$ exceeds roughly $0.5$ time units, well below $1/k_{AB}$.
In both sweeps, the MSM provides a reasonable but biased first approximation, and the RiteWeight iteration removes most of that bias.
Figure~\ref{fig:MB_Beta} tests the method as the inverse temperature $\beta$ is increased from 2 to 6, so that the barrier grows and transitions become exponentially rarer.
As discussed above, this is precisely the regime where brute-force simulation fails; the question is how the MSM approach scales.
We expect to pay for lower temperatures with more data, so the question is not whether $N$ must grow but how fast.
Two factors drive the per-trajectory variance of the sensitivity estimator up with $\beta$.
The score-function term in Eq.~\eqref{eq:DerivEM} carries an explicit factor of $\sqrt{\beta}$, contributing one factor of $\beta$ to the variance.
A second factor of $\beta$ comes from the increasing stiffness of the MSM transition matrix as the underlying timescales lengthen because the spectral gap between the stationary eigenvalue and the next-slowest mode shrinks, amplifying any error in the importance weights through the standard eigenvector-sensitivity relation~\cite{meyer_sensitivity_1994,thiede_sharp_2015}.
The same scaling has been observed in a reinforcement-learning context by \citet{cheng_surprising_2024}; analogous variance challenges arise more broadly in reinforcement-learning approaches to rare-event sampling, where they are addressed by learning control forces that bias the dynamics toward the events of interest~\cite{das_reinforcement_2021}.
The two combine to give a variance that grows as $\beta^2$, and we therefore scale $N = 5\times10^4\,\beta^2$ to hold the statistical error roughly constant across temperatures.
On that footing, any divergence in Fig.~\ref{fig:MB_Beta} reflects a coarse-graining error rather than a sample-size artifact.
We fix $M=20$ and $t=0.5$.
The RiteWeight sensitivity estimate tracks the exact reference closely throughout this range, while the MSM error grows with $\beta$.

\section{Optimizing the Bias of a Molecular Motor}
\label{sec:sec5}

The M\"{u}ller-Brown calculations tested the method against exact reference values in a two-dimensional landscape.
We now apply it to a problem where no such reference is available, optimizing the directional bias of a particle-based model of a chemically fueled molecular motor.
The motor model, introduced in \citet{albaugh_simulating_2022} and subsequently studied in Refs.~\cite{albaugh_sterically_2023,albaugh_precision_2024,penocchio_power_2025,gu_coupled_2025}, involves Langevin dynamics of dozens of interacting particles, a grand canonical Monte Carlo chemostat that drives the system out of equilibrium, and 30 adjustable Lennard-Jones parameters that control the interactions between the fuel and the catalytic machinery.
Computing sensitivities of the motor's bias to all 30 parameters simultaneously would be impractical by finite differences, but the MSM-based pipeline of Section~\ref{sec:sec2} yields the full gradient from a single ensemble of short trajectories.
Gradient descent on that gradient climbs the bias from about 50\% to above 90\% within roughly a dozen iterations, with parameter changes of at most tens of percent---a small, structured correction to a seemingly reasonable initial force field.

\subsection{Motor Model}
\label{sec:sec5a}

\begin{figure}[hbt]
\begin{center}
\includegraphics[width=.45\textwidth]{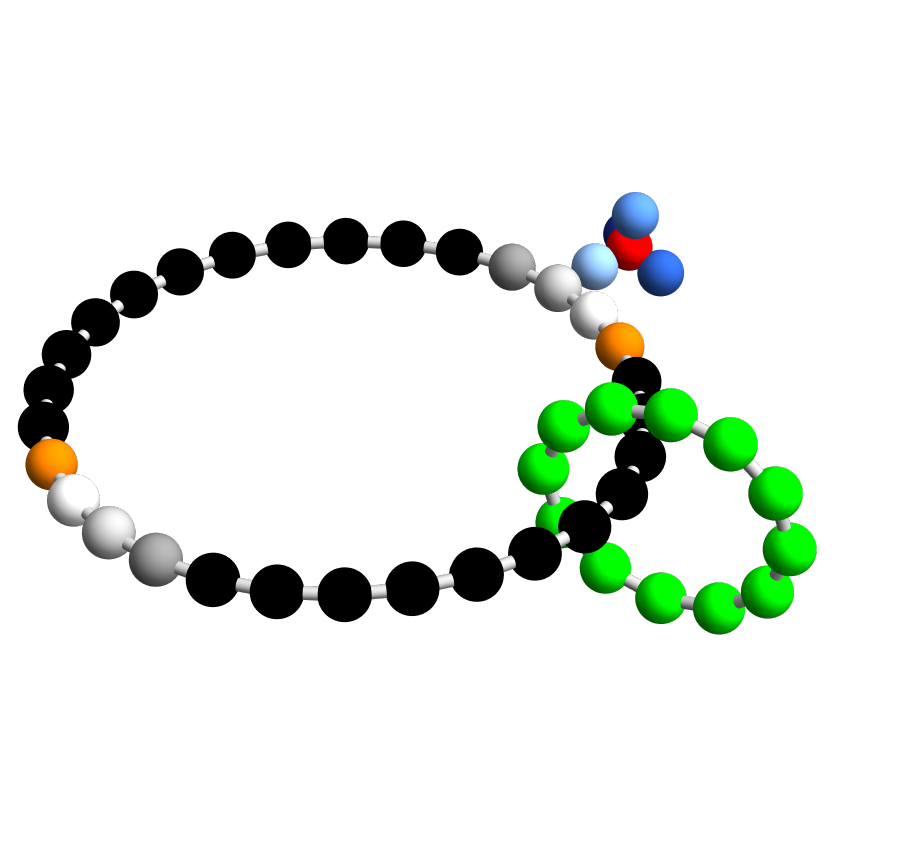}
\end{center}
\caption{\label{fig:MotorIllus}
Illustration of the catenane motor system.
The shuttling ring (purple) diffuses around the catalytic ring.
Binding sites (black beads) attract the shuttling ring.
Each three-bead catalytic site (green, orange, yellow) is adjacent to a binding site in the clockwise direction.
A tetrahedral fuel cluster (gray) carries a blocking group (blue) to the catalytic site, where interactions between the cluster and the catalytic beads catalyze release of the blocking group onto the orange bead.
Once deposited, the blocking group prevents the shuttling ring from passing.
}
\end{figure}

The motor is a [2]-catenane; it has two interlocked rings that cannot separate but can rotate relative to each other (Fig.~\ref{fig:MotorIllus}).
A large 30-bead ring serves as a track for a smaller 12-bead shuttling ring.
Most beads on the large ring are inert and purely volume-excluding.
Two binding-site beads, located at diametrically opposite positions, attract the shuttling ring, and adjacent to each binding site in the clockwise direction is a three-bead catalytic site (CAT1, CAT2, CAT3) that interacts with fuel molecules.
The fuel is a tetrahedral cluster of four particles (TET) enclosing a central blocking-group particle (C).
The intact cluster is called a full tetrahedral cluster (FTC); when the catalytic site stretches the cage and releases the blocking group, the remaining empty tetrahedral cluster (ETC) and free C particle dissociate.
Three grand canonical Monte Carlo chemostats maintain the populations of FTC, ETC, and C at prescribed chemical potentials, holding the system out of equilibrium with a thermodynamic driving force that favors FTC decomposition~\cite{albaugh_simulating_2022}.

Once deposited on the catalytic ring, the blocking group prevents the shuttling ring from passing.
Directionality arises because the shuttling ring, when bound to a binding site, sterically blocks fuel molecules from accessing the adjacent clockwise catalytic site while leaving the counterclockwise site accessible.
Blocking groups therefore accumulate preferentially in one direction, gating the shuttling ring's diffusion and producing a net current~\cite{albaugh_simulating_2022,albaugh_sterically_2023}.
The motor functions only within a narrow window of parameters.
The blocking group must dissociate neither too quickly (gating fails) nor too slowly (ring locks), and the shuttling ring's binding must be strong enough to suppress catalysis but not so strong as to immobilize it.
These competing requirements make gradient-based optimization a natural strategy.

The rings are held together by FENE bonds between adjacent beads and harmonic angle potentials that maintain circular geometry.
All pairwise interactions are described by a modified Lennard-Jones potential,
\begin{equation}
    V^{LJ}_{ij}(r_i,r_j)
    = 4\epsilon^r_{ij}\!\left(\frac{\sigma_{ij}}{|r_i-r_j|}\right)^{\!12}
    - 4\epsilon^a_{ij}\!\left(\frac{\sigma_{ij}}{|r_i-r_j|}\right)^{\!6},
\end{equation}
where $\sigma_{ij}$ is a pair-specific length scale and the repulsive and attractive amplitudes $\epsilon^r_{ij}$ and $\epsilon^a_{ij}$ are allowed to differ, reducing to the standard form when $\epsilon^a = \epsilon^r$; purely volume-excluding pairs have $\epsilon^a = 0$.
All bond, angle, mass, and Lennard-Jones radius parameters are identical to those in Ref.~\cite{albaugh_simulating_2022}; only the Lennard-Jones amplitudes governing fuel--catalytic-site interactions are varied in the optimization.
The starting values and simulation parameters are collected in Tables~\ref{tab:params} and~\ref{tab:paramsLJ}.

\begin{table}[ht]
\caption{\label{tab:params}%
Simulation and gradient-descent parameters for the molecular
motor optimization.
All quantities are in the non-dimensionalized units of
Ref.~\cite{albaugh_simulating_2022}.
The GCMC values $\mu'_i$ are the shifted chemical potentials $\mu_i - A^0_i$ defined in that reference, which absorb the cluster free energies and are the quantities that enter the Monte Carlo acceptance probabilities.
}
\begin{ruledtabular}
\begin{tabular}{lcc}
Parameter & Symbol & Value \\
\hline
\multicolumn{3}{l}{\textit{Gradient descent}} \\
Number of adjustable parameters  & ---        & 30   \\
Learning rate                     & $\eta$     & 0.007   \\
Maximum step cap                  & $c$        & 0.1   \\
\\
\multicolumn{3}{l}{\textit{MSM / trajectory data}} \\
Number of MSM coarse states       & $M$        & 531   \\
Lag time                          & $\tau$     & 0.1   \\
Trajectories per coarse state    & ---        & $10$ \\
Trajectory length                 & ---        & 80    \\
\\
\multicolumn{3}{l}{\textit{Langevin dynamics}} \\
MD timestep                       & $dt$       & $5\times10^{-3}$ \\
Inverse temperature               & $\beta$    & 2.0   \\
Damping coefficient               & $\gamma$   & 1.0   \\
\\
\multicolumn{3}{l}{\textit{GCMC chemical potentials}} \\
Filled tetrahedral cluster (FTC)  & $\mu'_\text{FTC}$  & 1    \\
Empty tetrahedral cluster (ETC)   & $\mu'_\text{ETC}$  & $-3$ \\
Central particle (C)              & $\mu'_\text{C}$    & $-10$ \\
\end{tabular}
\end{ruledtabular}
\end{table}

\begin{table}[ht]
\caption{\label{tab:paramsLJ}%
Starting Lennard-Jones parameters for the fuel--motor interactions that are optimized by gradient descent.
BG: blocking group (blue in Fig.~\ref{fig:MotorIllus});
CAT1, CAT2, CAT3: green, orange, yellow catalytic beads;
TET: gray tetrahedral cluster beads;
SHUTTLE: purple shuttling ring beads.  Parameters that are adjusted during the optimization are highlighted in blue.
}
\begin{ruledtabular}
\begin{tabular}{lcc}
Interaction & $\epsilon_r$ & $\epsilon_a$ \\
\hline
BG--CAT1             & \textcolor{blue}{1}       & \textcolor{blue}{0}    \\
BG--CAT3             & \textcolor{blue}{1}       & \textcolor{blue}{0}    \\
BG--CAT2             & \textcolor{blue}{3.0}     & \textcolor{blue}{4.0}  \\
BG--SHUTTLE          & 100000  & 0    \\
BG--all others       & 1       & 0    \\
TET(1--4)--CAT(1--3) & \textcolor{blue}{0.58}    & \textcolor{blue}{0.58} \\
TET(1--4)--all others & 1       & 0    \\
\end{tabular}
\end{ruledtabular}
\end{table}

\subsection{Directional Bias and Its Gradient}
\label{sec:sec5b}

The quantity we wish to maximize is the directional bias of the motor.
Let $N^\mathcal{T}_{\mathrm{CW}}$ and $N^\mathcal{T}_{\mathrm{CCW}}$ be the number of complete clockwise and counterclockwise cycles of the shuttling ring in a long interval $[0,\mathcal{T}]$.
The bias is
\begin{equation}\label{eq:BiasDef}
    B = \lim_{\mathcal{T}\to\infty}\frac{N^\mathcal{T}_{\mathrm{CW}}}{N^\mathcal{T}_{\mathrm{CW}}+N^\mathcal{T}_{\mathrm{CCW}}}.
\end{equation}
To write $B$ as a stationary expectation of the form~\eqref{eq:orderparam}, we introduce the discrete displacement $J_t$ of the shuttling ring.
Letting $j_t$ be the index of the track particle closest to the ring's center of mass at time $t$, that discrete displacement is given by
\begin{equation}\label{eq:DisplacementDef}
    J_{t+dt} = J_t + (j_{t+dt}-j_t) - 30\left\lfloor\frac{j_{t+dt}-j_t}{30}\right\rfloor.
\end{equation}
A complete clockwise (counterclockwise) cycle corresponds to $J$ crossing from 29 (from $-29$) back to 0.
If $\pi(X,J)$ is the joint stationary distribution of particle positions $X$ and displacement $J$, then
\begin{equation}
    \dot{N}_{\mathrm{(C)CW}} = \lim_{dt\to0}\frac{1}{dt}\left\langle\mathbbm{1}_{J_0=\pm29}\,\mathbbm{1}_{J_{dt}=0}\right\rangle_{\pi},
\end{equation}
and the bias $B = \dot{N}_{\mathrm{CW}}/(\dot{N}_{\mathrm{CW}}+\dot{N}_{\mathrm{CCW}})$ is a ratio of two expectations over $\pi$, each of the form treated in Eq.~\eqref{eq:orderparam}.
The structure is the same as the M\"{u}ller-Brown rate.
$B$ is a ratio of two $\langle F\rangle_\pi$ objects, so its gradient follows from the quotient rule applied to Eq.~\eqref{eq:ExpDeriv}, and the MSM importance weights and propagator derivatives are shared between numerator and denominator.

The parameters we optimize are 30 attractive and repulsive Lennard-Jones amplitudes governing fuel--catalytic-site and blocking-group--motor interactions (Table~\ref{tab:paramsLJ}).
Following Ref.~\cite{albaugh_simulating_2022}, the motor is confined to an inner simulation box by a wall potential, and GCMC insertions and deletions act only on a surrounding outer volume.
A freshly inserted or deleted molecule therefore has no direct interaction with the motor at the moment of the chemostat move, so the GCMC acceptance probability does not depend on the Lennard-Jones parameters being optimized.
Only the Langevin propagator derivative contributes to the bias gradient.
All other force-field parameters are held fixed at the values in Ref.~\cite{albaugh_simulating_2022}.

\subsection{Integrator and Propagator Derivative}
\label{sec:sec5a_integrator}

In the original motor simulations~\cite{albaugh_simulating_2022}, the dynamics were integrated with the Ath\`enes-Adjanor scheme \cite{athenes_measurement_2008}.
Here we switch to the ABOBA splitting \cite{leimkuhler_molecular_2015}, because ABOBA evaluates the force at a deterministic midpoint configuration rather than at a random one, which gives the propagator a closed Gaussian form and makes its log-derivative analytically tractable.
The integrator propagates both positions and momenta, but the friction coefficient ($\gamma = 1.0$) is large enough that inertial effects are negligible on the timescales relevant to the motor; the momentum degrees of freedom are retained because the ABOBA propagator has a convenient analytic form, not because the dynamics are inertially dominated.
Letting $X_t, P_t$ be the position and momentum, $\xi_t$ a standard normal random variable, $m$ the particle mass, $\gamma$ the friction coefficient, and $\nabla U_\lambda$ the force, the ABOBA half-steps are
\begin{align}
    X_{t+1/2} &= X_t + \tfrac{dt}{2m}P_t, \\
    P_{t+1/2} &= P_t - \tfrac{dt}{2}\nabla U_\lambda(X_{t+1/2}), \\
    \hat{P}_{t+1/2} &= e^{-\gamma dt}P_{t+1/2} + \sqrt{\beta^{-1} m (1-e^{-2\gamma dt})}\,\xi_t, \\
    P_{t+1} &= \hat{P}_{t+1/2} - \tfrac{dt}{2}\nabla U_\lambda(X_{t+1/2}), \\
    X_{t+1} &= X_{t+1/2} + \tfrac{dt}{2m}P_{t+1}.
\end{align}
Because the force is evaluated only at the non-random midpoint $X_{t+1/2}$, the propagator is Gaussian in $P_{t+1}$:
\begin{multline}\label{eq:PropagatorABOBA}
    P_{\lambda}(X_{t+1},P_{t+1}\mid X_t,P_t) = \\
    (2\pi f^2)^{-d/2}\exp\!\left[\frac{-(P_{t+1}-e^{-\gamma dt}P_t
    +G_t)^2}{2f^2}\right]\\
    \times\,\delta\!\left(X_{t+1}-X_t-\tfrac{dt}{2m}(P_{t+1}+P_t)\right),
\end{multline}
where $f = \sqrt{\beta^{-1}m(1-e^{-2\gamma dt})}$, $d$ is the dimension, and $G_t=\tfrac{dt}{2}(1+e^{-\gamma dt})\nabla U_\lambda(X_{t+1/2})$.
Taking the log derivative and recognizing the exponential argument as $\xi_t^T\xi_t/2$ yields
\begin{equation}\label{eq:DerivABOBA}
    \frac{d \log P_{\lambda}(X_{t+1},P_{t+1}\mid X_t,P_t)}{d\lambda}
    = -\sum_i \xi^i_t \,\frac{d(\nabla U_\lambda)^i(X_{t+1/2})}{d\lambda}\,g^i,
\end{equation}
where $g^i = dt(1+e^{-\gamma dt})/[2\sqrt{\beta^{-1}m^i(1-e^{-2\gamma dt})}]$, and $\xi^i_t$, $m^i$ are the noise sample and mass for particle~$i$ at step~$t$.
This gives an explicit, simulation-accessible expression for the propagator derivative in the second term of Eq.~\eqref{eq:ExpDeriv}.
The more common BAOAB splitting~\cite{leimkuhler_molecular_2015} would not admit such a direct expression because it evaluates the force at a random-input configuration, so its propagator does not have a closed Gaussian form in general.
We have verified that the ABOBA dynamics used here produce motor current and bias statistically consistent with those obtained from BAOAB, so the choice of splitting does not affect the physical conclusions.

\subsection{Collective Variables and MSM Construction}
\label{sec:sec5_msm}

The motor's configuration space is far too high-dimensional for a spatial Voronoi partition like the one used for M\"{u}ller-Brown.
Instead, we partition a small set of physically motivated collective variables into discrete MSM states.
The discretization is a triplet $(J_t, S^1_t, S^2_t)$.
The first index $J_t$ is the signed shuttling-ring displacement of Eq.~\eqref{eq:DisplacementDef}, which takes 59 integer values and tracks the slow progress of the ring around the track.
The second and third indices $S^i_t \in \{0, 1, 2\}$ report the state of each catalytic site.
$S^i_t = 2$ if a FTC sits within a distance $3.0$ of the central catalytic bead and $S^i_t = 1$ if no FTC is nearby but a blocking group lies within $2.6$ of that bead, and $S^i_t = 0$ otherwise.
Together these give $59 \times 3 \times 3 = 531$ MSM states, resolving both the shuttling ring's position and the gating state at each catalytic site.

The coarse transition matrix and importance-weight derivatives are computed exactly as in Section~\ref{sec:sec2a}, with Voronoi cell indicators replaced by indicators on these discrete states.
Because 531 states is still modest, the linear algebra in Eqs.~\eqref{eq:ChainRule}--\eqref{eq:PDerv} is inexpensive.
We use a single-iteration MSM for the motor without the RiteWeight refinement of Section~\ref{sec:sec3}.
The discrete labels $(J, S^1, S^2)$ do not come with a natural distance metric for generating candidate Voronoi centers, so the iterative reweighting would require a different construction of refined partitions than the one used for M\"{u}ller-Brown.
The agreement between the MSM-based gradient and direct simulation in the next section (Fig.~\ref{fig:motor_bias}) will confirm that a single MSM pass is adequate here.

\subsection{Optimization Protocol and Results}
\label{sec:sec5c}

Gradient descent requires an ensemble of initial configurations that covers the relevant MSM states.
At the first step we construct 10 copies of a configuration in each state, about 5310 configurations in total.
Each configuration is built by placing both rings as circles, rotating the shuttling ring to the position specified by $J$, inserting FTC and blocking-group particles to match the target $S^1, S^2$, and relaxing the result with 15{,}000 steps of energy minimization at step size $10^{-5}$.
Some MSM states place the shuttling ring directly on top of a fuel molecule or blocking group; we do not initialize configurations in these high-energy states, which leaves 451 accessible states at the first iteration.
Each configuration is then propagated for 80 time units with snapshots every 0.1 time units.
For subsequent gradient steps, starting configurations are drawn from the endpoints of the previous iteration's trajectories (10 per occupied MSM state).
Parameters are updated as
\begin{equation}
    \theta_{i+1} = \max\!\left\{0,\ \theta_i + \mathrm{clip}\!\bigl(\eta\,(\nabla_\theta B)_i,\, -c,\, c\bigr)\right\},
\end{equation}
where $\mathrm{clip}(x,-c,c) = \max\{-c,\min\{x,c\}\}$.
The cap $c$ prevents integrator instabilities and the outer $\max$ enforces non-negativity of the Lennard-Jones amplitudes.

To check that the MSM-based gradient reflects the underlying motor rather than a coarse-graining artifact, we also estimate the bias directly from 50 long trajectories of length $2\times 10^5$ time units at each parameter setting.
Each such trajectory spans roughly ten complete cycles, enough to determine the bias reliably.

\begin{figure}[hbt]
\begin{center}
\includegraphics[width=.45\textwidth]{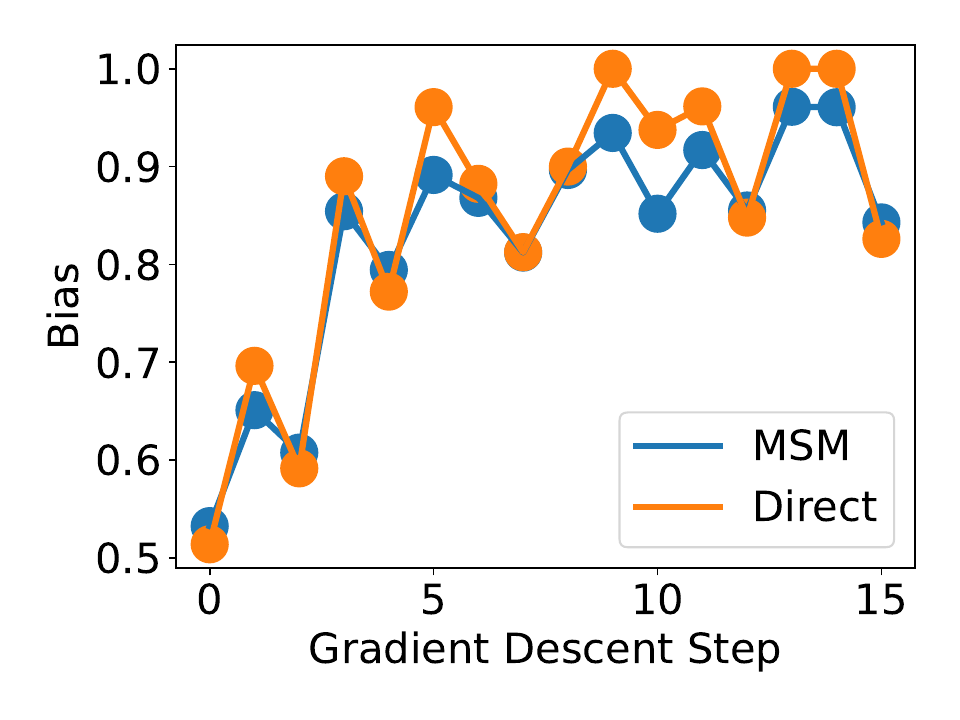}
\end{center}
\caption{\label{fig:motor_bias}
Motor directional bias $B$ as a function of gradient-descent iteration.
Blue circles: bias estimated from the MSM used to compute the gradient.
Orange squares: bias estimated independently from 50 long trajectories at each parameter setting.
The two estimates track each other, confirming that the gradient direction reflects the motor's actual bias rather than an artifact of the coarse graining.
}
\end{figure}

\begin{figure*}[hbt]
\begin{center}
\includegraphics[width=.9\textwidth]{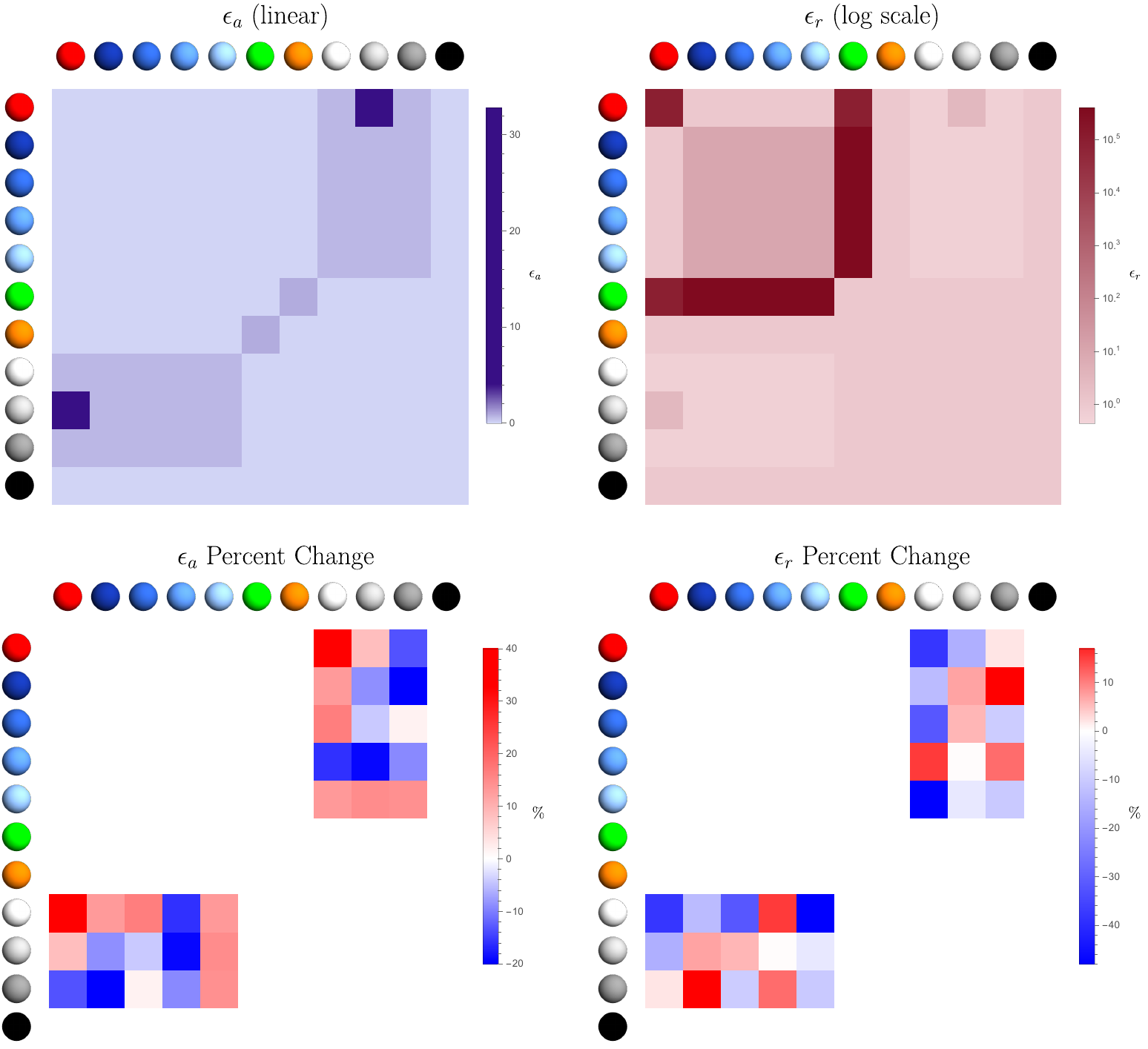}
\end{center}
\caption{\label{fig:motor_PctChange}
Top row: starting Lennard-Jones amplitudes; the repulsive parameters $\epsilon_r$ are shown on a log scale.
Bottom row: percent change in $\epsilon_a$ (left) and $\epsilon_r$ (right) between the starting force field and the 15th iteration.
The attractive amplitude $\epsilon_a$ between the fuel and the catalytic bead closest to the binding site (white-ball column) grows, while the amplitude between the fuel and the catalytic bead farthest from the binding site (dark-gray column) shrinks.
The repulsive amplitudes $\epsilon_r$ show the opposite trend.
The overall effect is to shift the catalytic event toward the binding site.
}
\end{figure*}

Figure~\ref{fig:motor_bias} shows the optimization in action.
The bias climbs from about 50\%---essentially no directional preference---to over 90\% within the first several iterations and then plateaus.
The MSM-based estimate used to compute the gradient and the independent long-trajectory estimate track each other throughout, so the gradient direction is not an artifact of the coarse graining.
Gradient descent works on this problem.

The more surprising finding in Fig.~\ref{fig:motor_PctChange} is that the transition from essentially unbiased to strongly biased motors requires modest nudges of the Lennard-Jones amplitudes.
Working and non-working motors, in other words, live close to each other in parameter space.
The same observation appears implicitly in the original motor model, where the ``Motor I'' and ``Motor II'' variants of Ref.\cite{albaugh_simulating_2022} differ by small parameter adjustments and yet show noticably different performance.
Our starting parameters were chosen to be a reasonable initial guess in that the necessary attraction between the blocking group and the catalytic site is set somewhat large, and the tetrahedral fuel is weakly attracted to the catalytic site.
Despite being a reasonable parameter set, the initial motor design generates no current because directional bias requires a specific pattern of attractions among fuel, catalytic site, and shuttling ring, and small perturbations of that pattern can turn a dud into a working motor.

The structure of the changes in Fig.~\ref{fig:motor_PctChange} reveals that pattern.
The attractive amplitudes $\epsilon_a$ grow between fuel particles and the catalytic bead closest to the binding site and shrink between fuel particles and the catalytic bead farthest from the binding site, with the opposite trend for the repulsive amplitudes $\epsilon_r$.
Gradient descent is shifting the catalytic event toward the binding site.
That shift amplifies the kinetic asymmetry underlying the motor's Brownian-ratchet mechanism~\cite{penocchio_power_2025,astumian_kinetic_2019}.
A shuttling ring resting on a binding site sterically suppresses catalysis on the adjacent clockwise site more effectively when that catalytic site is nearby, so biasing catalysis toward the near-side bead magnifies the gating difference between the two cycling directions.
The gradient has rediscovered, from a purely numerical criterion, the design principle argued for qualitatively in Refs.~\cite{albaugh_sterically_2023, penocchio_power_2025}.

Finally, a word on what we have and have not optimized.
Our objective is directional bias, not current or thermodynamic efficiency.
The three are distinct design targets, and they can trade off against each other.
A highly biased motor can turn slowly, and parameter changes that increase current may suppress bias~\cite{penocchio_power_2025}.
The pipeline accommodates any of these objectives with a different choice of $F$ in Eq.~\eqref{eq:orderparam}.
We selected bias because it has a clean form as a ratio of two counting expectations and so provides a convenient demonstration that gradient-based optimization can navigate a high-dimensional Lennard-Jones design space.

\section{Discussion}

We have presented a method for computing the sensitivity of dynamical order parameters to model parameters in systems with rare events.
The method works by separating the problem along the natural timescale separation.
Short trajectory data teaches a Markov state model how local rates depend on parameters, and linear algebra propagates those rate changes into changes in the stationary distribution.
The RiteWeight iteration systematically reduces the error introduced by the coarse-graining.

The central computational advantage is that we avoid long trajectories.
For problems dominated by rare events, the MSM importance weights allow us to represent the stationary distribution and its sensitivity from an ensemble of short trajectories, none of which need to span multiple barrier crossings.
The validation on the M\"{u}ller-Brown potential demonstrates that the method maintains accuracy even as the temperature is lowered and the timescale separation grows.

Beyond demonstrating the pipeline, the motor optimization sheds light on the parameter landscape of the \citet{albaugh_simulating_2022} model.
Seemingly dysfunctional choices of the Lennard-Jones amplitudes sit close in parameter space to working ones, and gradient descent on the bias systematically migrates catalysis toward the binding site---a mechanism consistent with the Brownian-ratchet account of \citet{penocchio_power_2025}.
The same pipeline would apply directly to any system whose dynamics can be described by Langevin dynamics or a similar explicitly parameterized propagator.

The strategy of differentiating through a coarse-grained kinetic model has appeared in other contexts.
\citet{trubiano_optimization_2022} combined MSM analysis with adjoint-based gradients to optimize a time-dependent protocol for maximizing the transient yield of a self-assembly target.
The key distinction is that their setting involves a time-varying protocol optimized for a finite-horizon objective, whereas here we compute steady-state sensitivities to static parameters.
Both approaches build an MSM and then differentiate through its transition matrix, suggesting that MSM-based gradient pipelines may be broadly useful for design problems in complex molecular systems, whether the objective is a steady-state property or a transient one.

Several limitations of the current approach deserve mention.
The method requires a coarse partition of state space, and the quality of the sensitivity estimate depends on how well that partition resolves the relevant dynamics.
For the M\"{u}ller-Brown potential, a Voronoi tessellation with a modest number of cells suffices; for higher-dimensional systems, choosing a good partition is itself a nontrivial problem, and adaptive or data-driven partitioning strategies \cite{husic_markov_2018} may be needed.
The ABOBA integrator was chosen because it yields an analytically tractable propagator derivative, but this choice is not a fundamental limitation of the framework.
Extending the method to integrators whose propagator lacks a closed form would require numerical differentiation or adjoint methods for the propagator derivative.
We have also worked throughout with observables that take the form of stationary expectations or ratios of such expectations; extending the pipeline to observables that do not (for instance, entropy production rates or thermodynamic efficiency) would require reworking the path functional $F$ in Eq.~\eqref{eq:ExpDeriv} but is conceptually straightforward.
Finally, the gradient descent in the motor application is not guaranteed to find a global optimum; the parameter landscape may contain local optima or saddle points that trap the optimization.

\section*{Code Availability}
The source code for the motor optimizations, as well as a user friendly jupyter notebook implementing the Muller Brown example are available in a Zenodo repository under the DOI:
https://doi.org/10.5281/zenodo.19950484

\begin{acknowledgments}
This work was initiated with support from the Gordon and Betty Moore Foundation under Grant Number GBMF10790.
This material is additionally based upon work supported by the U.S. Department of Energy, Office of Science, Office of Basic Energy Sciences program under Award Number DE-SC0026333.
\end{acknowledgments}

\bibliography{Motors}

\end{document}